\documentstyle[11pt,twoside,paspconf,epsf]{article}

\newcommand{\kms}{${\rm km \, s^{-1}}\,$}
\newcommand{\Ms}{${\rm M}_{\odot}\,$}

\begin{document}

\title{Detecting Halo Streams with GAIA}

\author{Amina Helmi, HongSheng Zhao and P. Tim de Zeeuw}
\affil{Sterrewacht Leiden, Postbus 9513, 2300 RA Leiden, The Netherlands}

\begin{abstract}
We investigate what the proposed ESA astrometric satellite GAIA will
reveal by observing the halo of the Milky Way. Specifically, we look
for halo streams which are the signatures left by the
merging/accretion events experienced by a typical galaxy like the
Milky Way. We run numerical simulations of the disruption of satellite
galaxies in a Galactic potential to generate artificial GAIA halo
catalogs.  We recover the streams by searching for peaks in angular
momentum space.
\end{abstract}


\section{Introduction}
GAIA is a short-listed candidate for an ESA cornerstone mission
provisionally scheduled for launch in 2009. It will provide precise
astrometry ($<$10$\mu$as in parallax and $<$10$\mu$as yr$^{-1}$ in
proper motion at $V \sim 15$, increasing to 0.2 mas yr$^{-1}$ at
$V\sim 20$) and multicolour photometry, for all 1.3 billion objects to
$V\sim20$, and radial velocities with accuracies of a few \kms for
most stars brighter than $V\sim17$, so that full six-dimensional
phase-space information will be available.  GAIA will thereby provide
a very large and statistically reliable sample of stars, from which
the questions concerning the origin and evolution of the Galaxy may
finally be answered.  In this contribution we address what GAIA may
tell us about the history and formation of the stellar halo of the
Milky Way.

Hierarchical theories of structure formation in the Universe propose
that galaxies are the result of mergers and accretion of smaller
building blocks. Such events would leave fossil signatures in the
present day components of the Milky Way, and in particular in its
stellar halo.  When a satellite galaxy is disrupted, it leaves trails
of stars along its orbit, so that when all accretion events are
superimposed, a spheroidal component may be produced.  Recent
observations have shown that indeed considerable structure is still
present in Milky Way's halo, indicating that accretion events have had
some role in its formation history (e.g. Majewski this volume).

There are several methods for detecting moving groups.  The Great
Circle Counts method (G3C) proposed by Johnston, Hernquist \& Bolte
(1996) uses the position on the sky, and employs the fact that
satellites in orbits that probe only the outer (spherical) halo
conserve the orientation of their plane of motion, thereby leaving
their debris along great circles on the sky, if observed from the
Galactic center (see also Johnston, this volume).  The methods used in
the Solar neighborhood for detection of disk moving groups and open
clusters use also proper motions (and sometimes parallax), and assume
that all the stars belonging to the same system have the same velocity
vector (e.g.\ Hoogerwerf \& Aguilar 1998; de Bruijne 1998).
Lynden-Bell \& Lynden-Bell's method (1995) needs the position on the
sky and the radial velocity, and has been used, for example, to link
globular clusters which lie on the same plane to some of the
(disrupted) dwarf companions of our Galaxy (see also Lynden-Bell, this
volume).

The applicability of the abovementioned methods is questionable in the
inner parts of the halo. In this regime, the Galactic potential is
strongly axisymmetric so that the debris does not remain on a fixed
plane, the situation where G3C works. As an example in Figure~1 we
show a sky projection of a satellite 8 Gyr after disruption: no strong
angular correlations are visible. On the other hand, even though the
velocity dispersions in a stellar stream do decrease with time, and
therefore, very strong correlations are to be expected (see Helmi \&
White this volume), in the inner halo strong phase-mixing takes place.
Since the superposition of streams can give a velocity dispersion as
high as 200 \kms, it would be rather difficult to detect satellite
debris if only velocity information is used.  Clearly, we need to
identify where the clustering that is characteristic of a satellite
manifests itself in the debris that we observe after many galactic
\nolinebreak orbits.\looseness=-2

\begin{figure}[thb]
\plotfiddle{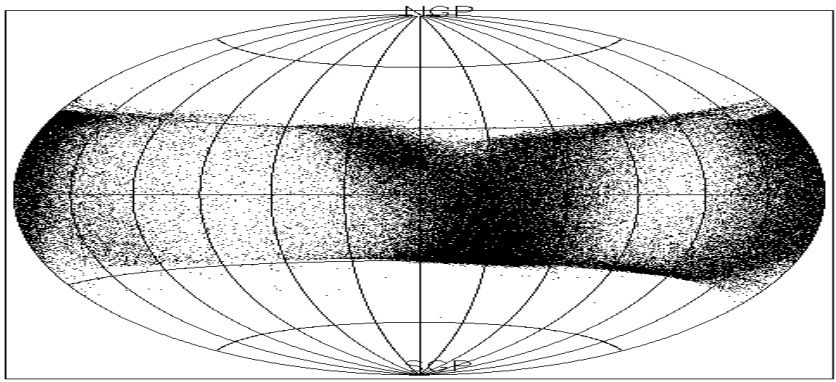}{60truemm}{0}{65}{27.3}{-160}{-20}
\caption{Galactocentric sky projection for one of our experiments.}
\end{figure}

A satellite can be considered an ensemble of particles with very
similar integrals of motion (energy, angular momentum) as shown in
Figure~2.  Since these are conserved quantities, or evolve only
slightly, this initial clumping should be present even after the
system has phase-mixed completely. Therefore, the space of integrals
or adiabatic invariants is the natural space to look for the
substructure produced by an accretion event.

\begin{figure}[!htb]
\plotfiddle{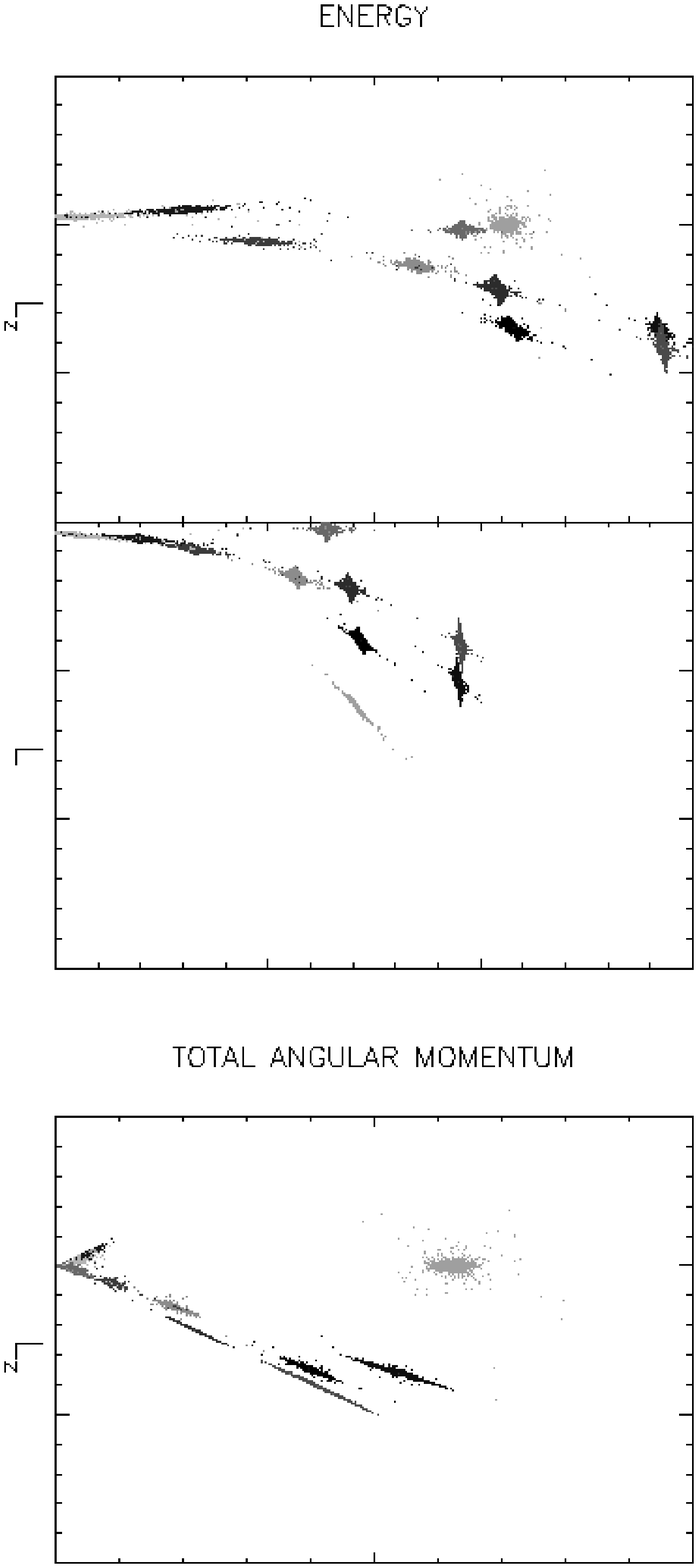}{30truemm}{90}{45}{30}{175}{-15} 
\caption{Initial distribution of particles for each of
our satellites in the space defined by ($E$, $L$, $L_z$). 
Each lump represents a different satellite. Only 2\% of the particles are
drawn.}
\end{figure}

We use angular momentum as a measure of the lumpiness in the stellar
halo. Even though it is not fully conserved for an axisymmetric
potential (only $L_z$ is), it evolves preserving a certain degree of
coherence. Since the computation of the angular momentum does not
involve any detailed knowledge of the Galactic potential, the method
is very powerful in identifying substructure if 6-D information is
available.  Moreover, the number of clumps detected in this way will
represent well the total number of accretion/merging events, since
unlike other methods which are only local, it singles out all the
stars from a given accreted object, independently of how different
their phases and velocities~might~be.

\section{Putting the method to work}

To generate an artificial GAIA catalog for the Galaxy we need to
include the accretion events, which are the substructure that we will
be searching for, and a smooth phase-space distribution of particles
for the disk and the bulge.  We describe here how to generate such a
data set, focusing on the accretion events and their detection. We
also discuss briefly the effect of the smooth component.

For the accreted component we use numerical simulations of the
disruption of satellite galaxies with masses $10^7 - 10^8$ \Ms,
initial dispersions $3-8$ \kms and sizes $1-2$ kpc. Their orbital
periods are in the range $0.5-1.5$ Gyr, and their pericentra lie in
the inner halo ($<$ 10 kpc).  We assume that the $10^5$ particles,
which represent each $10^7 - 10^8$ \Ms satellite, are KIII or MIII
stars, since this is roughly the expected number of giants of this
spectral type in a $10^9$ Gyr old object of such mass. This is to
consider only stars that are bright enough to be observable from the
Sun.  We convolve the positions and velocities of the particles
obtained from the simulations with the expected measurement errors,
given in Table~1.  If a star-particle is too faint to have a
measurable parallax it is left out of the analysis.  For a KIII star
this corresponds to $ V \sim 21$.

\begin{table}[b]
\caption{Estimated precision in parallax ($\sigma_\pi$, in $\mu$as)
and proper motion ($\sigma_\mu$, in $\mu$as ${\rm yr}^{-1}$) as a function
of V magnitude for a K3III star with no reddening (Lindegren 1998).
The precision in the radial velocity is taken to be 3 \kms, 
up to the magnitude limit.}
\begin{center}
\begin{tabular}{ccccccccccc}
\tableline 
& 10 & 13 & 14 & 15 & 16 & 17 &  18 & 19 & 20 & 21 \\
\tableline
$\sigma_\pi$ & 4.05 & 5.01 & 7.41 & 11.5 & 
18.2 & 29.9 & 51.6 & 96.8 & 202. & 609. \\
$\sigma_\mu$ & 2.43 & 3.01 & 4.44 & 6.86 & 
10.9 & 17.9 & 30.9 & 58.0 & 121. & 365. \\
\tableline 
\tableline 
\end{tabular}
\end{center}
\end{table}

\begin{figure}[thb]
\plotfiddle{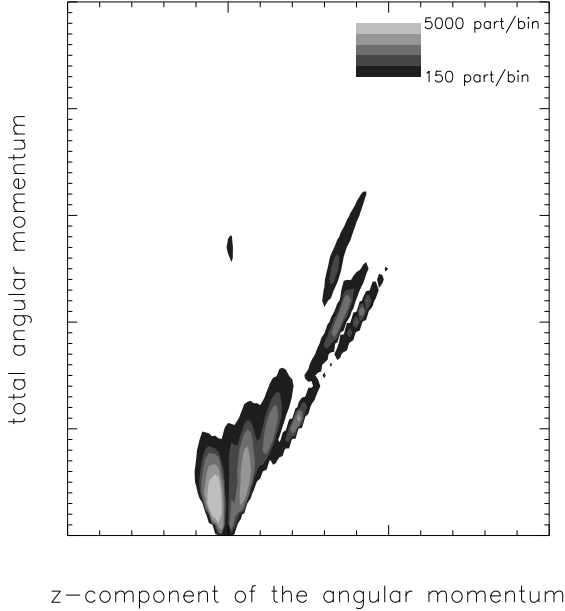}{75truemm}{0}{45}{45}{-150}{-70}
\caption{Density contours for the distribution of the `accreted' stars
in angular momentum space, after convolution with observational
errors, for our 10 satellites at 13.5 Gyrs after infall. We use a bin
of 100 kpc km/s in the $L_z$-direction and of 200 kpc km/s in the
$L$-direction. In this plot, the disk is located at $L = L_z$, and the
bulge dominates the region of $L < 2000$ kpc \kms.}
\end{figure}

Figure~3 shows the `observed' angular momentum distribution of the
particles that are left in the catalog after the previously described
analysis.  For comparison see the right panel of Figure~2 which
represents the initial distribution, prior to disruption and without
any measurement errors. The lumps remain coherent and hence the events
can be recovered.  The success of the method is largely due to the
full use of what characterizes a satellite in phase-space at all
times: its clumping in the adiabatic invariants space.  Secondly, the
accuracy of GAIA practically makes the initial and final distributions
(i.e. after error convolution) nearly indistinguishable.  It follows
that we should be able to easily detect any accreted satellite in the
high angular momentum part ($L > 2000$ kpc \kms) of the $L_z - L$
plane. In the lower regions of that diagram we may find superposition
of events, so that a more sophisticated analysis (not just the use of
contour plots) will be needed.  Moreover, the inclusion of the smooth
distribution due to the Galactic disk and bulge will also require a
deeper analysis, since the number of stars in these components is much
larger than that of the stellar halo.  For example, a large part of
the disk can be suppressed by removing from the catalog all stars
located within 1 kpc from the Galactic plane.  The disk can also be
identified easily in the $L_z - L$ plane as a narrow fringe located at
$L = L_z$. On the other hand, the bulge shares similar spatial and
kinematical properties with the stellar halo. In the $L_z - L$ plane
it is located in the $L < 2000$~kpc~\kms region. One way of
suppressing most of its contribution would be to separate bulge and
halo with a metallicity criterion (Minniti 1996). Nevertheless, we may
want to analyze the bulge in more detail, since hierarchical theories
predict that bulges may also form by accretion and merging events.

\section{Discussion}

From Figure~3 we conclude that the method is successful in detecting
the structure left by the accretion events.

One of the limitations of the method is the need of the 6-D
information which constrains the volume around the Galactic center
that we can probe with the GAIA measurements to $\sim$~20 kpc (in the
Sun's direction), however most of the halo stars are within that
volume. As discussed in the previous section, we may not be able to
resolve all accreted events. For example, if the number of disrupted
satellites is very large the chances of superposition will be larger.
However we can add extra dimensions using the conservation of energy
with a better constrained Galactic potential (see Zhao et al.\ this
volume).  We may also zoom in on those overlapping regions, to look for
smaller scale structure, and use as well velocity correlations.
	
The evolution of the Galactic potential may be the most crucial
simplification in our analysis.  In hierarchical cosmologies the
number of objects that form a galaxy like our own is in the range of
5-20, with comparable masses.  The process of formation is likely to
be very violent and the potential is surely not static, quite probably
not axisymmetric, and therefore the initial clumping of the system may
not be reflected in clumping in angular momentum space. However, if
this happened during the first few Gyrs, any object falling later,
ought to have perceived a fairly static (or adiabatically changing)
Galaxy, and then our method would still be of use.  This probably
covers masses up to several times $10^9$ \Ms.

\acknowledgments It is a pleasure to thank LKBF for financial support
and Mount Stromlo Observatory for hospitality. A.H.\ also acknowledges
Zonta International for the Amelia Earhart fellowship award.

\end{document}